\documentstyle[a4,12pt,axodraw]{article}

\def\title{A possible way to relate the ``covariantization'' and the 
negative dimensional integration method in the light-cone gauge}

\long\def\abstract{
In this work we present a possible way to relate the method
of covariantizing the gauge dependent pole and the negative
dimensional integration method for computing Feynman integrals
pertinent to the light-cone gauge fields. Both techniques are
applicable to the algebraic light-cone gauge and dispense with
prescriptions to treat the characteristic poles.\\
\\
{\bf PACS: 02.90.+p; 12.38.Bx}}

\def\fapesp{\  \epsfbox{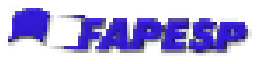} \  (Funda\c c\~ao de Amparo
                                \`a Pesquisa do Estado de S\~ao Paulo)}
\def\a{\alpha}\def\b{\beta}\def\g{\gamma}\def\o{\omega}\def\G{\Gamma}
\def\d{\delta}

\def\ndim{NDIM}
\def\NDIM{negative dimensional integration method}
\newcommand{\intq}{\int{d^{2\omega}q}\cdot}

\input{epsf}

\begin{document}

\vskip 15mm
\centerline{\huge\bf A possible way to relate the}
\centerline{\huge\bf ``covariantization'' and the negative}
\centerline{\huge\bf dimensional integration method}
\centerline{\huge\bf in the light-cone gauge}
\vskip 8mm
\begin{center} \vskip 10mm
Alfredo T. Suzuki{\begingroup\def\thefootnote{a}
                                \footnote{e-mail: suzuki@ift.unesp.br}
                                \addtocounter{footnote}{-1} \endgroup}
and \  \
Ricardo Bent\'{\i}n{\begingroup\def\thefootnote{z}
                                \footnote{e-mail: rbentin@ift.unesp.br}
                                \addtocounter{footnote}{-1} \endgroup}
\\[3mm] Instituto de F\'{\i}sica Te\'orica - UNESP.\\
        Rua Pamplona 145, CEP 01405-900, SP, Brasil.
        
\vskip 3.0cm              {\bf ABSTRACT } \end{center}    \abstract

\thispagestyle{empty} \newpage
\pagestyle{plain} 

\newpage
\setcounter{page}{1}

\section{\large Introduction.}

\baselineskip 1cm

Recently, Suzuki and Schmidt \cite{ndim} have calculated several one-
and two-loop Feynman integrals in the algebraic type gauges such as
the light-cone and the Coulomb ones using the negative dimensional
integration (NDIM) approach. For the former gauge, they have proven
that the \ndim{} approach can dispense with prescriptions and partial
fractionings as well as avoiding parametric integrals and integrations
over components. The only requirement in \ndim{} is the correct choice
of the light-like basis vectors that generate the four-dimensional
space-time. In this line, the second light-like ``dual'' vector
$m^\mu$, $(m^2=0)$, plays a key role in the whole process, otherwise
the calculations go wrong, violating causality and so on. Actually,
the gauge proper is defined with the one light-like vector, usually
denoted by $n^\mu$, $(n^2=0)$, such that $n\cdot A=0$, where $A_\mu$
is the gauge field; however, this condition is not enough. A residual
gauge freedom remains that must be eliminated via $m^\mu$ constraint.

The \ndim{} technique totally circumvents the need for prescriptions,
since the evaluation is done in negative dimension, or as understood,
as fermionic integrals (polynomial integrands) in positive dimension.
Here we would like to emphasize that the prescriptionlessness of the
\ndim{} approach in the light-cone gauge has nothing whatsoever to do
with the residual gauge freedom mentioned above. It has all to do with
the polynomial character of the fermionic integrands.

On the other hand, there is a technique coined by Suzuki \cite{cov} as
covariantization to treat light-cone integrals. This technique is
causal and reproduces the results obtained through the use of the
Mandelstam-Leibbrandt prescription \cite{ml}. This fact remains almost
forgotten, even though the covariantization manifestly guarantees the
absence of zero mode frequencies that spoil causality. Of course, here
the technique uses parametric integrations and integration over
components, and all the technology of complex analysis to treat the
singularities in a proper way. The thrust of the technique lies in
that it does not require a prescription for the light-cone pole; it
``converts'' this pole in a ``covariant'' pole whose treatment is
well-grounded and established since the early days of quantum field
theory. The burden of this technique and its most severe drawback is
that it requires an additional parametric integration to be performed,
a task which can be very demanding.

Since we are going to use the light-cone gauge throughout this work,
we will briefly describe the light-cone coordinates. Using the metric
$(+,-,-,-)$, then a general contravariant four-vector is given by:
$$
  x^{\mu}=(x^+,x^-,x^i) \ ,   \ i=1,2,
$$
where:
\begin{eqnarray*}
    x^{\pm} &=& {\frac{1}{\sqrt{2}}}(x^0{\pm}x^3), \\
                &=&{\frac{1}{\sqrt{2}}}(x_0{\mp}x_3)=x_{\mp},\\
  x^i&{\equiv}&(\hat{x})^i =(x^1,x^2)
\end{eqnarray*}
Now, defining the light-like four-vectors:
\begin{eqnarray}
  \nonumber
  n_{\mu} &=& {\frac{1}{\sqrt{2}}}(1,0,0,1),\\
  \label{lcc}
  m_{\mu} &=& {\frac{1}{\sqrt{2}}}(1,0,0,-1),
\end{eqnarray}
we observe that
\begin{eqnarray*}
        x^+ &=& x^{\mu}n_{\mu},\\
        x^- &=& x^{\mu}m_{\mu},
\end{eqnarray*}
and the scalar product becomes:
$$ 
x^{\mu}y_{\mu}=x^+y^-+x^-y^+-{\hat{x}}{\hat{y}},
$$
i.e.,
$$ x^2=2x^+x^--{\hat{x}}^2$$.

This means that the metric is now:
\begin{eqnarray*}
  g^{{\mu}{\nu}}&=&       \begin{array}{cc}
                                \begin{array}{cccc}
                                ^+ & ^- & ^1 & ^2
                                \end{array}
                                & \\
                                \left(
                                        \begin{array}{cccc}
                                                0 & 1 & 0 & 0 \\
                                                1 & 0 & 0 & 0 \\
                                                0 & 0 & 1 & 0 \\
                                                0 & 0 & 0 & 1\\
                                        \end{array}
                                \right)
                                &
                                \begin{array}{c}
                                ^+ \\ ^- \\ ^1 \\ ^2
                                \end{array}
                                \end{array}
\end{eqnarray*}

\section{\large What is meant by covariantization}

Here is a brief review of the ``covariantization'' technique which was
proposed by Suzuki \cite{cov}. The idea is quite simple. In light-cone
coordinates, the square of a four-momentum is:
$$
  q^2=2q^+q^--\hat{q}^2,  
$$
As long as {\it $q^-\ne0$}, we can write $q^+$ as
$$
  q^+=\frac{q^2+\hat{q}^2}{2q^-}
$$

We note that this dispersion relation almost guarantees that real
gauge fields for which $q^2=0$ (real photons or real gluons for
example) are transverse; the residual gauge freedom, that is left to
be dealt with so that fields be manifestly transverse comes from the
presence of the $q^-$ in the denominator of the expression above.

This implies that in the light-cone gauge the characteristic pole 
becomes
\begin{equation}
  \frac{1}{q^+}=\frac{2q^-}{q^2+\hat{q}^2},
\end{equation}
The important thing here is that the condition $q^-\ne0$ warranties
the causal structure of this technique since it eliminates the
troublesome $q^-=0$ modes. Elimination of these modes restores the
physically acceptable results as can manifestly be seen in the causal
prescription \cite{pz} for the light-cone gauge.

\section{\large Negative dimension integration.}

This method of integrating Feynman loop integrals is a follow up
development of the early attempts by Halliday, and Ricotta \cite{hall}
in this direction during the mid 80's. In the late 90's it was tested
in the case of working in the light-cone gauge \cite{us}.  The best
way to illustrate this method is through an example. So we will use
the computation of the following integral:
$$
  I_{ijk}=\intq(q-a)^{2i}(q-b)^{2j}(q^+)^k.
$$
Note that for $i=j=k=-1$ this integral is the basic one-loop integral
in the light-cone gauge. In \ndim{}, we start with this integral
considering {\em positive} values for the exponents $i,j$, and $k$ and
{\em negative} dimension $D=2\omega$ (or as state above, as a
fermionic polynomial integration in positive dimension). First of all,
let us define the generating Gaussian functional for the integral in
question, that is,
$$
  I_{Gauss}= \intq
  e^{\scriptscriptstyle -\a(q-a)^2-\b(q-b)^2-\g q^+}
$$
which clearly can be rewritten as
\begin{equation}
  \label{trick1}
  I_{Gauss}=
  \sum^\infty_{ijk=0}(-)^{i+j+k}
  \frac{\scriptstyle \a^i+\b^j+\g^k}
  {i!j!k!}\overbrace{\intq (q-a)^{2i}(q-b)^{2j}(q^+)}^{I_{ijk}}.
\end{equation}

On the other hand, performing the momentum integration, which is not
difficult to do since it is a Gaussian integral, we arrive at
$$
  I_{Gauss}=\left(\frac{\pi}{\a+\b}\right)^\o
  e^{-\frac{1}{\a+\b}[\a\b(a-b)^2-\a\g a^+ \b\g b^+]}.
$$
Expanding the last exponential and also using the multinomial 
expansion, we obtain
\begin{equation}
  \label{trick2}
  I_{Gauss}=\pi^\o\sum^\infty_{n_i=0}(-)^{n_{123}}
  \frac{\a^{n_{124}}\b^{n_{135}}\g^{n_{23}}}{n_1!n_2!n_3!n_4!n_5!}
  (-n_{123}-\o)!(a-b)^{n_1}(a^+)^{n_2}(b^+)^{n_3},
\end{equation}
where $i=1\cdots 5$, $n_{abc}=n_a+n_b+n_c$. Comparing now Eq. 
(\ref{trick1})
and Eq. (\ref{trick2}) we conclude that $I_{ijk}$ is given by:
\begin{eqnarray*}
  I_{ijk}&=&\pi^\o(-)^{i+j+k}\G(1+i)\G(1+j)\G(1+k)
  \sum^\infty_{n_i=0}(-)^{n_{123}}
  \frac{\a^{n_{124}}\b^{n_{135}}\g^{n_{23}}}{n_1!n_2!n_3!n_4!n_5!}
  (-n_{123}-\o)!\\
  &&\cdot(a-b)^{n_1}(a^+)^{n_2}(b^+)^{n_3}\cdot
  \d_{n_{124},i}\d_{n_{135},j}\d_{n_{23},k}\d_{n_{45},-n_{123}-\o}.
\end{eqnarray*}
The $\d$'s form a system of equations which in this case have four
constraints, 
\begin{eqnarray*}
  n_1+n_2+n_4&=&i,\\
  n_1+n_3+n_5&=&j,\\
  n_2+n_3&=&k,\\
  n_1+n_2+n_3+n_4+n_5+\o&=&0.\\
\end{eqnarray*}

Clearly the usual matrix equation does not close and no solution can
be found since there are five unknowns and only four equations. The
system can only be solved when we pretend one of the unknows is a
known variable. Of course, there is no preferred unknown to play this
role; we can choose it in several ways.  In fact, in the present case,
there are $C^5_4$ ways to do this, i.e., {\it five solutions}.  One
such solution is a trivial one, and then we have two sets of pairs of
solutions (each pair belonging to a peculiar region of external
momenta). And the general solution for the pertinent Feynman integral
is constructed as a linear combination of the particular solutions we
have for the system, according to the outcoming variables defined by
the external momenta.

Each general solution will have as a rule two factors: a coefficient
factor of gamma functions which we express in terms of Pochhammer's
symbols and eventually multi-indexed series which is typically
generalized hypergeometric functions of some sort, whose parameters
depends on the exponents of propagators and dimension as well as the
external momenta and vectors $n$ and $m$. The next step is to perform
an analytic continuation of exponents into the negative values and
positive dimension via a property of Pochhammer's symbols, such that
we obtain for our present example,
\begin{eqnarray*}
  I_{ijk} &=& {\pi}^{\o}[(a-b)^2]^{\rho-k}\left\{(b^+)^k 
    A \;{}_2F_1(-k,{\o}+j;1+j-\rho|z)\right\}\\
  && +\left(a^+)^{\rho-j}(b^+)^{-i-\o} 
    C \;{}_2F_1({\o}+i,{\rho}+{\o};1+{\rho}-j|z)\right\},
\end{eqnarray*}
where
\begin{eqnarray*}
  A&=&(-j|{\rho})(-i|i+j+{\o})({\rho}+{\o}|-2{\rho}-{\o}+k),\\
  C&=&(-j|i+j+{\o})(-i|i+k-{\rho})(-k|j+k-{\rho}),      
\end{eqnarray*}
with $z=\frac {a\cdot n}{b\cdot n},$ and $\rho = i+j+k+\o$.
Also, the Pochhammer symbols is defined as:
$$
  (a|b)=(a)_b=\frac{\G(a+b)}{\G(a)},
$$
with the property: $(a|b+c)=(a+b|c)(a|b)$.
Taking the limit $i,j,k\rightarrow -1$ (i.e. {\it positive dimension}
now) and using a property of the hypergeometric functions, it is 
possible to write it as just one term
\begin{eqnarray*}
  I_{-1-1-1}={\pi}^{\o}{\frac{[(a-b)^2]^{{\omega}-2}}{a^+}}
        {\G}(2-{\o})B({\o}-1,{\o}-1)\ _2F_1(1,{\o}-1;2{\o}-2|u).
\end{eqnarray*}

Now, if we also make the limit $a\rightarrow 0$, the result is in
complete agreement with that one found in Ref.\cite{capper}, but it is
a result in the context of the PV prescription for the light-cone
gauge, which is wrong. We use the above computation in order to show
how the negative integration method works. Now we are going to obtain
the right answer as it was done in Ref.\cite{us}. To do this, first
note that when we defined the light-cone coordinates in Eq.\ref{lcc},
there exists the dual of $n_\mu$, which we called $m_\mu$, and in
light-cone computations, this vector {\it always} appears in the {\it
numerator} of the integrands. Keeping this in mind, we will compute
the following integral using \ndim{} technology:
$$
  T_{ijk}=\intq[(q-p)^2]^i\:(q^+)^j\:(q^-)^k,\:\: k\ge 0.
$$
Using the same procedure as above, we get
$$
  T_{ijk}=(-\pi)^{\o}\left(\frac{-2p^+p^-}{n\cdot m}\right)^{i+\o}
  (p^+)^j (p^-)^k\frac{(1-i-\o|2i+\o)}{(1+j|i+\o)(1+k|i+\o)},
$$
Note that since from the start we must have $k\ge 0$ the Pochhammer
symbol containing the exponent $k$ , namely, $(1+k|i+\o)$ cannot be
analytic continued and the final result, after analytic continuation
to positive dimension and negative values of exponents $i$ and $j$ is
\begin{equation}
  \label{res}
  T_{ijk}=\pi^{\o}\left(\frac{2p^+p^-}{n\cdot m}\right)^{i+\omega}
  (p^+)^j(p^-)^k\frac{(-j|-i-\o)}{(i+\o|-2i-\o)(1+k|i+\o)}
\end{equation}
This result is in complete agreement with those obtained via 
causal prescriptions, such as the Mandelstam-Leibbrandt, but in 
this case without the use of any prescriptions. 

\section{\large How they are related.}

Negative dimensional integration method is a prescriptionless method
in the light-cone gauge. So it is the ``covariantization''
technique. Then, the natural question to ask is whether they are
related to each other in any way. Let us consider for definiteness and
simplicity but without loss of generality, the general form of the
one-loop light-cone integral,
\begin{equation}
  \label{start}
  F(p)=\intq\frac{f(q,p,n)}{g(q,p)(q^+)^\g},\ \ \ \ \g>0,
\end{equation}
where $f(q,p,n)$ and $g(q,p)$ are some well behaved functions.  Also,
let $\g=\a+\b$, with $\a$ and $\b$ being non-negative numbers.  This
means that we have two sets of choices for $\a$ and $\b$. The first
one is $\a\ge 0$ and $\b>0$. The second one is $\a>0$ and
$\b\ge0$. Then Eq.(\ref{start}) will be,
$$
  F(p)=\intq\frac{f(q,p,n)}{g(q,p)(q^+)^\a(q^+)^\b}.
$$
In our analysis, we can choose whatever case. It is indifferent which
one we take, since actually we can interchange $\a\leftrightarrow\b$
and the conclusion does not change. Therefore we take $\a>0$ and
$\b\ge 0$ for definiteness.  Use of the ``covariantization'' method
means that for the $\b$-pole we have,
$$
  \frac{1}{(q^+)^\b}=\frac{2^\b(q^-)^\b}{(q^2+\hat{q}^2)^\b},
$$
Since we have chosen the case where $\a$ is strictly positive, $\a>0$,
we keep this $\a$-pole as the caracteristic light-cone pole, and then
\begin{equation}
   \label{here}
   F(p)=\intq\frac{f(q,p,n)2^\b(q^-)^\b}{g(q,p)
     (q^2+\hat{q}^2)^\b(q^+)^\a}
     =\intq\frac{F(q,p,n)(q^-)^\b}{G(q,p)(q^+)^\a},
\end{equation}
where we have defined $F(q,p,n)=2^\b f(q,p,n)$ and
$G(q,p)=g(q,p)(q^2+\hat{q}^2)^\b$. Since $\b\ge 0$ and $\a>0$, we can
see that this is {\it the very condition} we used to define the
prescriptionless \ndim{} integral in {\it negative dimensions}. To
illustrate these manipulations, take the following case:
$f(q,p,n)=2^\b (q^2-\hat{q}^2)^{-\b}$ and $g(q,p)=(q-p)^2$, so that
Eq.(\ref{here}) takes the form:
\begin{equation}
   F(p)=\intq\frac{(q^-)^\b}{(q-p)^2(q^+)^\a}
\end{equation}
which is nothing more than the starting integral $T_{ijk}$ for the
\NDIM{}. Remember that after the analytic continuation was made, we
had to take the values of $\a=-1$ and $\b=0$ and put it in
Eq.(\ref{res}).  On the other hand, if we use the identities of the
``covariantization'' technique, the integral of Eq.(\ref{here}) we
arrive is,
$$
  F(p)=\intq \frac{1}{(q-p)^2(q^+)^{\a+\b}},
$$
but $\a+\b=1$ and with these values the above integral yields (using 
the covariantization) \cite{cov}
\begin{equation}
  \label{ult}
  F(p)=i(-\pi)^\o(2p^+p^-)^{\o-1}\frac{\G(2-\o)\G(\o-1)}{\G(\o)},
\end{equation}
If we manipulate the result obtained in Eq.(\ref{res}) (computed using
negative integration) and compare with the above Eq.(\ref{ult})
(computed using ``covariantization''), we will find that they are the
same solution.

\section{\large Conclusions.}

Here we have shown that the two prescriptionless ways to treat the
spurious poles of the light-cone gauge, are in fact related to each
other. The results in both cases are in concordance with those ones in
which some causal prescription is used. It is quite clear from the
start that the covariantization method is done in positive dimensions
whereas the \ndim{} is done in negative dimensions. What we were
interested in was to see the possibility of relating one with the
other and find some common grounds for both techniques. What we have
found is that the covariantization technique, which ensures the
troublesome $q^-=0$ modes to be left out of the computation in a
natural way can be seen as the crucial factor from which to start with
in negative dimensions. In other words, since $q^-\ne 0$ throughtout
any computation to be done, one sees that the four-vector $m^\mu$
cannot be left out of the general structure in \ndim{} applied to the
light-cone gauge.

\vspace{.5cm}

{\bf Acknowledgments:} 

One of us (R.B.) would like to thank {\fapesp} for financial support.


\begin{thebibliography}{11}
\bibitem{ndim} A. Suzuki, A.G.M. Schmidt, Eur. Phys. J. C 5 
  (1998) 175.\\
  A. Suzuki, A.G.M. Schmidt, Can. J. Phys. 78 (2000) 769.\\
  A. Suzuki, A.G.M. Schmidt, J. Phys. A 33 (2000) 3713.\\
  A. Suzuki, A.G.M. Schmidt, Phys. Lett. B 494 (2000) 332.
\bibitem{cov}  A.T. Suzuki, Mod. Phys. Lett. A 
  8 25 (1993) 2365.  
\bibitem{ml} S. Mandelstam, Nucl. Phys. B 213 (1983) 149.\\
  G. Leibbrantd, Phys. Rev. D 29 (1984) 1699.
\bibitem{pz} B.M. Pimentel and A.T. Suzuki, Phys. Rev. D 42
  (1990) 2115.
\bibitem{hall} I.G. Halliday and R. Ricotta, Phys. Lett. B 193
  (1987) 241.\\
  G.V. Dunne and I.G. Halliday, Phys Lett. B 193 
  (1987) 247.
\bibitem{us} A. Suzuki, A. Schmidt, R Bent\'{\i}n,
  Nucl. Phys. B 537 (1999) 549.
\bibitem{capper} D.M. Capper, J.J. Dulwichand M.J. Litvak, Nucl. Phys.
  B 241 (1984) 463.
\end{thebibliography}
\end{document}